\documentclass[prd,superscriptaddress]{revtex4-2}
\usepackage{amsmath,amssymb}
\usepackage{latexsym}
\newcommand{\be}{\begin{equation}}
\newcommand{\ee}{\end{equation}}
\begin{document}

\title{Generalized MICZ-Kepler systems on three-dimensional sphere and hyperboloid}

\author{Levon Mardoyan}
\email{mardoyan@ysu.am}
\affiliation{Bogoliubov Laboratory of Theoretical Physics, Joint Institute for Nuclear Research, Dubna, Russia}
\affiliation{Yerevan State University, 1 Alex Manoogian St., Yerevan, 0025, Armenia}

\author{Armen Nersessian}
\email{arnerses@yerphi.am}
\affiliation{Yerevan Physics Institute, 2 Alikhanian Brothers St., Yerevan 0036 Armenia}
\affiliation{Institute of Radiophysics and Electronics, Ashtarak-2, 0203, Armenia}
\affiliation{Bogoliubov Laboratory of Theoretical Physics, Joint Institute for Nuclear Research, Dubna, Russia}

\begin{abstract}
We propose analogs of the generalized MICZ-Kepler system on the three-dimensional sphere and the two-sheeted hyperboloid. We construct their energy spectra and normalized wave functions and find that they depend on two quantum numbers, which suggests that the systems are minimally superintegrable.
\end{abstract}
\maketitle

\section{Introduction}
About five decades ago, Zwanziger \cite{Zw} and McIntosh and Cisneros \cite{MIC} independently suggested a maximally superintegrable generalization of the three-dimensional Kepler-Coulomb system, specified by the presence of a Dirac monopole and inheriting the qualitative properties of the Kepler-Coulomb problem. This system is now known as the MICZ-Kepler system. The similarity between the MICZ-Kepler and Kepler-Coulomb systems has a transparent explanation. As is known, the Kepler-Coulomb system can be obtained from the four-dimensional oscillator via the so-called Kustaanheimo-Stiefel transformation assuming reduction by a $U(1)$ group action \cite{KS}. For zero value of the $U(1)$ generator this transformation yields the Kepler-Coulomb system, while for a non-zero value it yields the MICZ-Kepler system \cite{iwai}.

The Kepler-Coulomb system admits a generalization to spheres and pseudospheres (two-sheted hyperboloids) while preserving its maximal superintegrability property \cite{Higgs}.
The (pseudo)spherical generalizations of the MICZ-Kepler system that inherit maximal superintegrability are also known \cite{NP}.
Moreover, any classical/quantum mechanical system with a central potential defined on a three-dimensional $so(3)$-invariant space admits a ``MICZ-Kepler extension'' preserving its qualitative properties.
For this purpose one should incorporate the Dirac monopole and modify the potential term as follows \cite{mny}:
\be
V(r)\quad \rightarrow\quad V(r)+\frac{ \hbar^2s^2}{2\mu g(r)r^2},
\label{1}\ee
where $\mathbf{r}=(x_1,x_2,x_3)$ are the coordinates in which the metric takes the conformally flat form
\be
ds^2=g(r)d\mathbf{r} d\mathbf{r},
\label{cf}
\ee
$s=0,\pm 1/2, \pm 1,\ldots $ is the Dirac monopole number, and $\mu$ is the mass of the probe particle.

Textbook examples of integrable deformations of the Kepler-Coulomb problem, e.g., the two-center Kepler-Coulomb problem and the Stark-Coulomb problem, also admit MICZ-extensions and further (pseudo)spherical generalizations \cite{KNO,Oz,Vahagn}.
In contrast with these systems, the study of ``MICZ-extensions'' of less known (``non-textbook'') integrable deformations of the Kepler-Coulomb system is in its initial stage.

An exception is the so-called ``generalized Kepler-Coulomb system'' defined by the potential
\be
V_{gCoulomb}= \frac{\lambda_{1}}{r(r+x_3)} +\frac{\lambda_{2}}{r(r-x_3)}-
\frac{e^{2}}{r}, \qquad{\rm where}\qquad r=|\mathbf{r}|,\quad \mathbf{r}=(x_1,x_2,x_3),\quad \lambda_1,\lambda_2\geq 0.
\label{gCoulomb}
\ee
This system has four independent constants of motion (and thus belongs to the class of ``minimally superintegrable systems'') \cite{smor} and has been studied in great detail both classically and quantum mechanically (see, e.g., \cite{evans}).
For $\lambda_1=\lambda_2\neq 0$, the potential \eqref{gCoulomb} reduces to the Hartmann potential \cite{Hartmann}, which has been used for describing axially symmetric systems like ring-shaped molecules and studied from various viewpoints \cite{Hothers}.

The ``MICZ-extension'' of the generalized Kepler-Coulomb system, now known as the ``generalized MICZ-Kepler system'', was suggested by one of the authors in \cite{Mardoyan1} and subsequently studied in further publications, see, e.g., \cite{gMICZ}.

However, its generalizations to curved spaces have not yet been considered.

In this paper we fill this gap by proposing the ``generalized MICZ-Kepler systems on the three-dimensional sphere and hyperboloid (pseudosphere)''. In stereographic coordinates, where the metric takes the conformally flat form \eqref{cf}, these systems are defined by the potential
\be
 V_{gMICZ}=\frac{1}{g(r)} \left(\frac{\hbar^2s^2}{2\mu r^2}+
 \frac{\lambda_{1}}{r(r+x_3)} +\frac{\lambda_{2}}{r(r-x_3)}\right)+V(r)
,\label{gCoulombSphere}\ee
where
\be
 g(r)= \left(1+\frac{\varepsilon r^2}{4R^2_0}\right)^{-2},\qquad V(r)=-\left(1- \frac{\varepsilon r^2}{4R^2_0}\right)
\frac{e^2}{r} +\frac{1-\varepsilon}{2}\frac{e^2}{R_0},\qquad \varepsilon=\pm 1.
\label{gV}\ee
The case $\varepsilon=1$ corresponds to the system on the three-dimensional sphere, and $\varepsilon=-1$ to the three-dimensional hyperboloid (pseudosphere).
Respectively, $r\in[0,\infty)$ for the sphere, and $r\in[0, 2R_0)$ for the hyperboloid, with $R_0$ being their radii.

We construct the spectra and normalized wavefunctions of these systems and find that their spectra depend on two quantum numbers, i.e., they should have four functionally independent integrals (including the Hamiltonian) and be minimally superintegrable.

Furthermore, from our consideration it becomes clear that the expression \eqref{gCoulombSphere}
defines the relevant ``generalized MICZ-extension'' not only for the Coulomb system on the three-dimensional (pseudo)sphere,
but for a particle moving on any $so(3)$-invariant space (equipped with the metric \eqref{cf}) in the presence of any central potential $V(r)$. Hence, it can be considered as the generalization of the correspondence given by \eqref{1}.\\

The paper is organized as follows.\\
In \textbf{Section 2} we briefly discuss the generalized (Euclidean) MICZ-Kepler system \cite{Mardoyan1} and present expressions for its spectrum and wavefunctions, which will be used in further considerations.\\
In \textbf{Sections 3 and 4} we propose the generalized MICZ-Kepler systems on the three-dimensional sphere and pseudosphere (upper sheet of the two-sheeted hyperboloid) and find their spectra and normalized wavefunctions.\\
In \textbf{Section 5} we summarize the obtained results and discuss possible future developments.

\section{Generalized MICZ-Kepler system}
In this Section we present the main properties of the generalized MICZ-Kepler system, which will be used in our future considerations.
This system is defined by the Hamiltonian \cite{Mardoyan1}
\be
\widehat{\mathcal{H}}_{gMICZ}=\frac{\mathbf{\widehat{p}}^2}{2\mu}+\frac{\hbar^2s^2}{2\mu r^2}+\frac{\lambda_1}{r(r+x_3)}+\frac{\lambda_2}{r(r-x_3)}-\frac{e^2}{r},
\label{cgMICZ}
\ee
where
\be
\widehat{\mathbf{p}}:=-\imath\hbar\boldsymbol{\partial}-\frac{e}{c}\mathbf{A},\quad  \mathbf{A}=\frac{g}{r(r-x_3)}\left(x_2,-x_1,0\right)\;:\quad \left[\widehat{p}_i,\widehat{p}_j\right]=-\imath \hbar s\frac{\varepsilon_{ijk}x^k}{r^3},
\quad \left[\widehat{p}_i,x_j\right]=-\imath\hbar\delta_{ij}.
\label{p}\ee
Clearly, $\mathbf{A}$ is the vector potential of a Dirac monopole with magnetic charge $g = \hbar c s/e$ $(s = 0, \pm 1/2, \pm 1, \dots)$, and with the singularity axis $x_3 > 0$ \cite{Dirac}.

It should be noted that the Schr\"odinger equation for the generalized MIC-Kepler system admits separation of variables not only in spherical and parabolic coordinates, but also in prolate spheroidal coordinates \cite{Mardoyan1}.
Moreover, the generalized MIC-Kepler system can be obtained from the four-dimensional double singular oscillator via the Kustaanheimo-Stiefel transformation \cite{Mardoyan3}.

For $\lambda_{1,2}=0$, the generalized MICZ-Kepler system reduces to the original MICZ-Kepler system, which possesses conserved quantities given by the angular momentum and Runge-Lenz vector:
\be
\widehat{\mathbf{J}}= \frac{1}{\hbar}\mathbf{r}\times\widehat{\mathbf{p}} -s\frac{\mathbf{r}}{r},\qquad \widehat{\mathbf{I}}=\frac{1}{2\mu}\left( \widehat{\mathbf{J}}\times\widehat{\mathbf{p}}-\widehat{\mathbf{p}}\times\widehat{\mathbf{J}}\right)-\frac{{e^2  }}{\hbar}\frac{\mathbf{r}}{r}.
\label{JI}
\ee
Clearly, these operators are not constants of motion for the generalized MICZ-Kepler system due to the perturbing term in the Hamiltonian \eqref{cgMICZ}.
Instead, the generalized MICZ-Kepler system has the following constants of motion:
\be
\widehat{J}_3= x_{2}\widehat{p}_{1}-x_{1}\widehat{p}_{2}-\frac{sx_{3}}{r}, \qquad
\widehat{M}^s=\widehat{\mathbf{J}}^2 +\frac{2\lambda_1 r}{r+x_3}+\frac{2\lambda_2r }{r-x_3},\qquad \widehat{\cal I}\equiv \widehat{I}_3+\frac{\lambda_1(r-x_3)}{r(r+x_{3})}-\frac{\lambda_2(r+x_3)}{r(r-x_3)} :
\label{cInt}
\ee
\be
[\widehat{{\cal H}}_{gMICZ},\widehat{J}_3]=[\widehat{{\cal H}}_{gMICZ},\widehat{M}^s]=[\widehat{{\cal H}}_{gMICZ},\widehat{{\cal I }}]=0,
\qquad
[\widehat{J}_3,\widehat{M}^s]=[\widehat{J}_3,\widehat{{\cal I}}]=0,\qquad [\widehat{{\cal I}},\widehat{M}^s]=\widehat{S},
\ee
where $\widehat{S}^2$ is a cubic function of the operators \eqref{cInt}.
Hence, the generalized MICZ-Kepler system is a minimally superintegrable three-dimensional system: it has four functionally independent constants of motion (including the Hamiltonian), with three of them being in involution.
Therefore, its energy spectrum should depend on two quantum numbers. Below we will see that this is indeed the case.

Let us consider the spectral problem:
\be
\widehat{H}_{gMICZ}\psi =E\psi,\quad \widehat{M}^s\psi =\widetilde{j}(\widetilde{j}+1)\psi,\quad \widehat{J}_3\psi=m\psi.
\label{sp}
\ee
Following \cite{Mardoyan1} (see also \cite{Mardoyan4}), we introduce spherical coordinates
$x_1+\imath x_2=r\sin\theta\; {\rm e}^{\imath\varphi}, x_3=r\cos\theta$.

In these terms the first two commuting constants of motion in \eqref{cInt} depend only on angular coordinates:
\begin{eqnarray}
&&\widehat{J}_3=    -\imath \frac{\partial }{\partial\varphi }+s,\label{eq170}\\
&& \widehat{M}^{(s)} = -\left\{\frac{1}{\sin\theta}\frac{\partial}{\partial \theta}
\left(\sin\theta \frac{\partial}{\partial \theta}\right) + \frac{1}{4\cos^{2}\frac{\theta}{2}}
\left(\frac{\partial^{2}}{\partial \varphi^{2}} - \frac{4\mu\lambda_1}{\hbar^2} \right) +
\frac{1}{4\sin^{2}\frac{\theta}{2}}\left[\left(\frac{\partial}{\partial \varphi} + 2\imath s\right)^{2}
- \frac{4\mu\lambda_2}{\hbar^2} \right]\right\}.\label{eq17}
\end{eqnarray}
Then we immediately find that the spectral problem \eqref{sp} admits separation of variables in these coordinates:
\begin{eqnarray}
\psi \equiv \psi_{n j m}^{(s)}\left(r, \theta, \varphi\right) =
R_{n j m}^{(s)}\left(r\right)\,
Z^{(s)}_{jm}\left(\theta, \varphi\right),
\label{eq8}
\end{eqnarray}
where the angular wave function becomes an eigenfunction of the commuting operators $\widehat{M}^{(s)}$, $\widehat{J}_{3}$:
\be
\widehat{M}^{(s)}\,Z^{(s)}_{jm}\left(\theta, \varphi \right) =\widetilde{j} (\widetilde{j}+1)
Z^{(s)}_{jm}\left(\theta, \varphi\right),
\qquad
\widehat{J}_{z}\,Z^{(s)}_{jm}\left(\theta, \varphi\right) =
m\,Z^{(s)}_{jm}\left(\theta, \varphi\right).
\label{eq16}
\ee
The eigenvalues in the spectral problem \eqref{sp} are defined by the expressions
\begin{eqnarray}
&&E_{n,m}^{(s)} = - \frac{\mu e ^{4}}{2\hbar^{2}\left(n +\delta_{m}^{(s)} \right)^{2}},\qquad n = |s|+1, |s|+2, \ldots,\label{16}\\ 
&&\widetilde{j} = j+\delta_{m}^{(s)},\qquad j=m_{+}, m_{+}+1,\ldots, n - 1,\qquad  m=-j, -j+1, \ldots, j-1, j.\label{17}
\end{eqnarray}
Here and in what follows we use the notation
\be
m_+ = \frac{\left|m+s\right| + \left|m-s\right|}{2},\quad \delta_{m}^{(s)}=\frac{ m_1+m_2}{2}-m_+,\quad
m_1=  \sqrt{\left(m-s\right)^{2}+\frac{4\mu\lambda_{1}}{\hbar^{2}}},\quad m_2= \sqrt{\left(m+s \right)^2 +\frac{4\mu\lambda_{2}}{\hbar^{2}}  }.
\ee

The wavefunctions are given by the expressions
\be
R^{(s)}_{nj}= \frac{2\kappa^{2}\sqrt{r_{0}}}
{\Gamma(2j+2 \delta_{m}^{(s)} +2)}\,
\sqrt{\frac{\Gamma(n + j + 2\delta_{m}^{(s)}+ 1)}{(n - j -1)!}}
\left(2\kappa r\right)^{j+\delta_{m}^{(s)}}
e^{-\kappa r}  F(-n + j + 1; 2j+2\delta_{m}^{(s)} +2; 2\kappa r),
\label{eq9}\ee
\be
Z^{(s)}_{jm}
=\sqrt{\frac{(2j+2\delta_{m}^{(s)} +1)(j-m_{+})!
\Gamma(j+m_1+m_2-m_+ +1)}
{4\pi \Gamma(j +m_{1}- m_{+}+1) \Gamma(j +m_2- m_{+}+1)}}\left(\cos\frac{\theta}{2}\right)^{m_{1}}
\left(\sin\frac{\theta}{2}\right)^{m_{2}} P^{(m_{2},m_{1})}_{j-m_{+}}\left(\cos\theta\right)
\,e^{\imath (m-s)\varphi},
\label{eq10}
\ee
where
$r_{0} = \hbar^{2}/\mu e^{2}$ is the Bohr radius, $\kappa = \sqrt{-2\mu E}/\hbar$, $F\left(a; c; x\right)$ is the confluent hypergeometric function, and $P^{(a,b)}_{n}(x)$ is the Jacobi polynomial.
 Let us   notice that the single-valuedness of wavefunction requires to $m, j$ be (half-)integer  for   (half-)integer   $s$.

With these expressions at hand we are ready to construct and solve the quantum-mechanical generalized MICZ-Kepler systems on the three-dimensional sphere and hyperboloid.

\section{GENERALIZED MICZ-KEPLER SYSTEM ON THE THREE-DIMENSIONAL SPHERE}
The three-dimensional sphere of radius $R_0$ embedded in the four-dimensional Euclidean space $\mathbf{R}^4$ equipped with the metric $d\mathbf{y} d\mathbf{y} +dy_4 dy_4$ is defined by the constraint
$\mathbf{y}^2+y_4^2=R^2_0$, where $\mathbf{y}=(y_1,y_2, y_3)$.
This constraint can be solved in stereographic  coordinates $\mathbf{r}=(x_1,x_2,x_3)$:
\be
\mathbf{y}= \frac{ \mathbf{r}}{1+\frac{\mathbf{r}^2}{4R^2_0}}, \quad \mathbf{y}_4= R_0\frac{1-\frac{\mathbf{r}^2}{4R^2_0} }{1+\frac{\mathbf{r}^2}{4R^2_0} }
\ee
and  in hyperspherical coordinates:
\begin{eqnarray}
y_{1} + \imath y_{2} = R_0\sin\chi \sin\theta e^{\imath\varphi}, \qquad y_{3} = R_0\sin\chi \cos\theta, \qquad y_{4} = R_0\cos\chi, \qquad   \chi, \theta \in [0, \pi], \quad\varphi \in [0, 2\pi).
\label{eq2-1}
\end{eqnarray}
These coordinates are related as follows:
\begin{eqnarray}
x_{1} + \imath x_{2} = \frac{2R_0 \sin\chi \sin\theta e^{\imath\varphi}}{1+\cos\chi}, \qquad x_{3} =\frac{2R_0\sin\chi \cos\theta }{1+\cos\chi}.
\label{eq2-1p}
\end{eqnarray}
The induced metric on the sphere then takes the form
\be
ds^2=\left(d\mathbf{y} d\mathbf{y} +dy_4 dy_4\right)\vert_{\mathbf{y}^2+y_4^2=R^2_0}=\frac{d\mathbf{r}d\mathbf{r}}{(1+\frac{\mathbf{r}^2}{4R_0^2})^2}=R^2_0\left(d\chi^2+\sin^2\chi(d\theta^2 +\sin^2\theta d\varphi^2)\right).
\ee
Thus, in stereographic  coordinates the metric takes the conformally flat form \eqref{cf} with
\be
g(r)= \left(1+\frac{\mathbf{r}^2}{4R_0^2}\right)^{-2}.
\label{gS}\ee
The Coulomb potential on the three-dimensional sphere is defined as a harmonic potential, i.e., a solution of the Laplace equation on the three-dimensional sphere, and has the form \cite{Schrod}:
\begin{eqnarray}
V (\chi) =-\frac{e^2}{R_0}\frac{y_4}{|\mathbf{y}|} =- \left({1-\frac{\mathbf{r}^2}{4R_0^2}}\right) \frac{e^2}{r}=-\frac{e^{2}}{R_0}\cot\chi,
\label{eq2-2}
\end{eqnarray}
while its dual is just the vector potential of a Dirac monopole (see, e.g., the second reference in \cite{KNO}).

We then define the ``generalized spherical MICZ-Kepler analog'' of the generalized Kepler-Coulomb potential \eqref{gCoulomb} as follows:
\be
V_{gMICZ} = \frac{1}{g(r)}\left(\frac{\hbar^2s^2}{2\mu r^2} + \frac{\lambda_1}{r(r+x_3)}+ \frac{\lambda_2}{r(r-x_3)}\right)  - \left({1-\frac{\mathbf{r}^2}{4R_0^2}}\right) \frac{e^2}{r},
\ee
where $g(r)$ is defined by the expression \eqref{gS}. 

The Hamiltonian then takes the form
\be
\widehat{\mathcal{H}}_{gMICZ}= -\frac{\hbar^2}{2\mu}\triangle_{s}+V_{gMICZ}=-\frac{\hbar^2}{2\mu} \triangle_r +
\frac{\hat{\mathbf{J}}^2-\hbar^2s^2}{2\mu g(r)\, r^2}+V_{gMICZ}
\label{hamilt}
\ee
where
\be
  -\hbar^2\triangle_{s}\equiv \frac{1}{\sqrt{{\det g}}}\hat{p}_i g^{ik}\sqrt{\det
g}\;\hat{p}_k,\qquad \triangle_r \equiv
\frac{1}{g^{3/2}r^2}\frac{\partial}{\partial
r}\left(g^{1/2}r^2\frac{\partial}{\partial r}\right).
\label{hamiltR}\ee
The momentum generators are defined by \eqref{p} and include the potential of the Dirac monopole, and 
the rotational momentum generators $\mathbf{J}$ are as in the flat case \eqref{JI}.

The Schr\"odinger equation for the generalized MICZ-Kepler problem on the three-dimensional sphere can be written in hyperspherical coordinates as follows:
\begin{eqnarray}
\left\{\frac{1}{\sin^{2}\chi}\frac{\partial}{\partial \chi}
\left(\sin^{2}\chi \frac{\partial}{\partial \chi}\right) - \frac{1}{\sin^2\chi}\widehat{M}^{(s)} + \frac{2\mu e^{2}R_0}{\hbar^{2}}\cot\chi +  \frac{2\mu R^{2}_0}{\hbar^{2}}E \right\}
\Psi(\chi) = 0,
\label{eq2-3}
\end{eqnarray}
where the operator $\widehat{M}^{(s)}$ is given by \eqref{eq17}.
Hence, it admits separation of variables in hyperspherical coordinates. Namely, choosing the wavefunction in the form
\begin{eqnarray*}
\Psi(\chi,\theta,\varphi) = \mathcal{R}(\chi) Z^{(s)}_{jm}\left(\theta, \varphi \right)
\end{eqnarray*}
and taking into account the equation \eqref{eq16}, we obtain the following quasi-radial Schr\"odinger equation
\begin{eqnarray}
\frac{1}{\sin^{2}\chi}\frac{d}{d \chi}
\left(\sin^{2}\chi \frac{d \mathcal{R} }{d \chi}\right) +
\Biggl[\frac{2\mu R^{2}_0}{\hbar^{2}}E - \frac{ \widetilde{j} (\widetilde{j} +1 )}{\sin^{2}\chi}
 + \frac{2\mu e^{2}R_0}{\hbar^{2}}\cot\chi\Biggr]
\mathcal{R} (\chi ) = 0. \nonumber
\label{eq2-4} \end{eqnarray}
Its solution is given by the following energy spectrum (see  \cite{KMP})
\be 
E_{n,m} = \frac{\hbar^{2}}{2\mu R^{2}_0}\left[\left(n +\delta_{m}^{(s)} \right)^{2}
- 1\right] - \frac{\mu e^{4}}{2\hbar^{2}\left(n + \delta_{m}^{(s)}  \right)^{2}},\qquad n = |s|+1, |s|+2, \ldots,
\label{eq2-10}
\ee
and by the quasi-radial wavefunction
\begin{eqnarray}
\mathcal{R}_{nj}(\chi) = C_{nj}^{(s)}
\left(\sin\chi\right)^{j + \delta_{m}^{(s)}}
{\rm e}^{-\imath\chi(n - j - 1 - \imath\sigma)}
{_{2}F}_{1}\left(-n + j + 1, j +\delta_{m}^{(s)}  + 1 + \imath\sigma;
2j +2\delta_{m}^{(s)}   + 2; 1 - e^{2\imath\chi}\right),
\label{eq2-11} \end{eqnarray}
with the normalization constant $C_{nj}^{(s)}$ given by the expression (for the details of its derivation see the subsection below)
\begin{eqnarray}
C_{nj}^{(s)} = \frac{2^{j + 1 +\delta_{m}^{(s)}}}
{\Gamma\left(2j +2\delta_{m}^{(s)}  + 2\right)}e^{\pi \sigma/2}
\left|\Gamma\left(j + \delta_{m}^{(s)}   + 1 - \imath\sigma\right)\right|
\sqrt{\frac{\left[\left(n +\delta_{m}^{(s)}  \right)^{2} + \sigma^{2}\right]
\Gamma\left(n + j +2\delta_{m}^{(s)}  + 1\right)}
{\pi R^{3}_0\left(2n + 2\delta_{m}^{(s)}\right)\left(n - j - 1\right)!}},
\label{normsphere}\end{eqnarray}
where 
\be
\sigma =  \frac{R_0}{r_0\left(n+\delta^{(s)}_m\right)} = 
\frac{\mu e^2 R_0}{\hbar^2\left (n+\delta^{(s)}_m\right)}.
\label{sigma}\ee
The range of validity of $j,m$ is defined by \eqref{17}.

The reality of the quasi-radial wave function (\ref{eq2-11}) can be verified using the relation \cite{Bateman1}:
\begin{eqnarray*}
{_{2}F_{1}}\left(a, b; c; z\right) = \left(1 - z\right)^{-a}{_{2}F_{1}}\left(a, c - b; c; \frac{z}{z - 1}\right).
\end{eqnarray*}
In the limit $s = \lambda_{1} = \lambda_{2} = 0$, expressions \eqref{eq2-10}, \eqref{eq2-11}, \eqref{normsphere} reduce to the corresponding formulas for the hydrogen atom on the three-dimensional sphere, obtained in \cite{Schrod,Stiv,Mardoyan5}.\\

It is important to notice that the constructed spectrum and wavefunctions depend on only two quantum numbers, indicating that the generalized MICZ-Kepler system on the three-dimensional sphere could/should have an additional integral generalizing the third expression in \eqref{cInt}.  

\subsection{Calculation of normalization constants}
Let us compute the expression for the normalization constant \eqref{normsphere} for the generalized MICZ-Kepler system on the three-dimensional sphere.

The normalization condition for the quasi-radial wave function of this system reads:
\begin{eqnarray}
R^{3}_0\int\limits_{0}^{\pi}\sin^{2}\chi \mathcal{R}_{n'j}^{*}(\chi ) \mathcal{R}_{nj}(\chi ) d\chi = \delta_{nn'}.
\label{eq2-12}
\end{eqnarray}
Since the range of the argument of the hypergeometric function appearing in \eqref{eq2-11} on the real axis lies on the segment $[0, 2]$, it is convenient to pass to a hypergeometric function of the argument $e^{2\imath\chi}$, whose range on the real axis lies on the segment $[0, 1)$. Using the known relation for the hypergeometric function \cite{Bateman1}:
\begin{eqnarray}
_{2}F_{1}(a, b; c; z) = \frac{\Gamma(c)\Gamma(c - a - b)}
{\Gamma(c - a) \Gamma(c - b)} {_{2}F}_{1} (a, b; a + b + 1 - c; 1 - z )
+ \nonumber \\
\label{eq2-13} \\
+ \frac{\Gamma(c)\Gamma(a + b - c)}
{\Gamma(a) \Gamma(b)}(1 - z)^{c - a - b} {_{2}F}_{1}(c - a, c - b;
c+ 1 - a - b; 1 - z), \nonumber
\end{eqnarray}
we can represent the quasi-radial wave function \eqref{eq2-11} in the following form:
\begin{eqnarray}
&\mathcal{R}_{nj}(\chi)& = C_{nj}^{(s)} \frac{\Gamma (2j + 2\delta_{m}^{(s)}     + 2 )
\Gamma (n +  \delta_{m}^{(s)}  - \imath\sigma )}
{\Gamma (n + j +2\delta_{m}^{(s)}   +1 )
\Gamma (j + \delta_{m}^{(s)} + 1 - \imath\sigma )}  (\sin\chi )^{j +\delta_{m}^{(s)}}
{\rm e}^{-\imath\chi (n - j - 1 - \imath\sigma )}  \nonumber
 \\ [4mm]
&&\times
{_{2}F}_{1} (-n + j + 1, j + \delta_{m}^{(s)}+ 1 + \imath\sigma;
1 - n -\delta_{m}^{(s)} + \imath\sigma; e^{2\imath\chi} ).
\label{eq2-14}\end{eqnarray}
Now, substituting the functions \eqref{eq2-14} into \eqref{eq2-12}, and representing the hypergeometric functions as polynomials, we perform the integration in accordance with the following relations \cite{Bateman1}:
\begin{eqnarray}
\int\limits_{0}^{\pi} \left(\sin t\right)^{\alpha} e^{\imath\beta t} dt= \frac{\pi \Gamma\left(1 + \alpha\right)
e^{\imath \pi \beta/2}}{2^{\alpha} \Gamma\left(1 + \frac{\alpha + \beta}{2}\right)
\Gamma\left(1 + \frac{\alpha - \beta}{2}\right)}\quad{\rm when}\quad \Re e\;\alpha > 0,\qquad
\frac{\Gamma\left(z\right)}{\Gamma\left(z - n\right)} = \left(-1\right)^{n}
\frac{\Gamma\left(-z + n + 1\right)}{\Gamma\left(-z + 1\right)}.
\label{eq2-15}
\end{eqnarray}

Then for $n' = n$ we get:
\begin{eqnarray}
\frac{\pi R^{3}_0e^{-\pi \sigma}\left|C_{nj}^{(s)}\right|^{2}}{2^{2j +2\delta_{m}^{(s)}  + 2}}
\Gamma (2j +2\delta_{m}^{(s)}   + 3 )
\Biggl[
\frac{\Gamma ( 2j +2\delta_{m}^{(s)}  + 2 )\left|\Gamma (n +\delta_{m}^{(s)}   -\imath\sigma )\right|}
{\Gamma  (n + j +2\delta_{m}^{(s)}  + 1 )\left|\Gamma (j + \delta_{m}^{(s)}   + 1 - \imath\sigma )\right| }  \Biggr]^{2}\times\nonumber \\ [2mm]
\times
\sum\limits_{p=0}^{n-j-1}\frac{ (-n + j + 1 )_{p}
 (j +\delta_{m}^{(s)}     + 1 + \imath\sigma )_{p}}
{p!  (1 - n - \delta_{m}^{(s)}    + \imath\sigma  )_{p}}
 \frac{\left(-1\right)^{p}}{\Gamma (j +\delta_{m}^{(s)}   + p + 2 + \imath\sigma )
\Gamma (j + \delta_{m}^{(s)}   - p + 2 - \imath\sigma )}\times \nonumber \\ [2mm]
\times{_3F_2}\left\{\begin{matrix} -n + j + 1,\,\,j +  \delta_{m}^{(s)}  + 1 - \imath\sigma,\,\, -j - \delta_{m}^{(s)}    - p - 1 - \imath\sigma\\
1 - n - \delta_{m}^{(s)} - \imath\sigma,\,\, j +    \delta_{m}^{(s)}  - p + 2 - \imath\sigma \end{matrix}\Biggr|1\right\}
= 1 \; ,
\label{eq2-16}\end{eqnarray}
 where $(a)_m=\Gamma(a+m)/\Gamma(a)$ is the Pochhammer symbol.

Now, using twice the relation \cite{Bailey}:
\begin{eqnarray}
{_3F_2}\left\{\begin{matrix} a,\,\,a',\,\, -N\\
b',\,\, 1-N-b \end{matrix}\Biggr|1\right\}=\frac{(a+b)_{N}}{(b)_{N}}\,
{_3F_2}\left\{\begin{matrix} a,\,\,b'-a',\,\, -N\\
b',\,\, a+b \end{matrix}\Biggr|1\right\}
\label{eq2-17}
\end{eqnarray}
we obtain:
\begin{eqnarray*}
&& {_3F_2}\left\{\begin{matrix} -n + j + 1,\,\,j +  \delta_{m}^{(s)}    + 1 - \imath\sigma,\,\, -j - \delta_{m}^{(s)}  - p - 1 - \imath\sigma\\
1 - n - \delta_{m}^{(s)}   - \imath\sigma,\,\, j + \delta_{m}^{(s)}    - p + 2 - \imath\sigma \end{matrix}\Biggr|1\right\} = \\ [2mm]
&=& \frac{\left(-p\right)_{n-j-1}(-n - j - 2 \delta_{m}^{(s)}   - 1)_{n-j-1}}
{(-n - \delta_{m}^{(s)}    + p - \imath\sigma)_{n-j-1}
(j +\delta_{m}^{(s)}  + 1 + \imath\sigma)_{n-j-1}}\;
 {_3F_2}\left\{\begin{matrix} -n + j + 1,\,\, -j - \delta_{m}^{(s)}   - p - 1 - \imath \sigma,\,\, - 1\\
- n - j - 2 \delta_{m}^{(s)}  - 1,\,\,  - p  \end{matrix}\Biggr|1\right\}.
\end{eqnarray*}
Further, taking into account that according to the formula \cite{Bateman1}:
\begin{eqnarray}
\frac{\Gamma\left(-z + n\right)}{\Gamma\left(-z\right)} = \left(-1\right)^{n}
\frac{\Gamma\left(z + 1\right)}{\Gamma\left(z - n + 1\right)}.
\label{eq2-17-1}
\end{eqnarray}
one has:
\begin{eqnarray*}
(-n - j - 2\delta_{m}^{(s)}    - 1)_{n-j-1} =
(-1)^{n-j-1}
\frac{\Gamma(n + j + 2\delta_{m}^{(s)}   + 2)}
{\Gamma(2j + 2\delta_{m}^{(s)}   + 3)},
 \end{eqnarray*}
 \begin{eqnarray*}
 (-n - \delta_{m}^{(s)}    + p + \imath\sigma )_{n-j-1} =
 (-1 )^{n-j-1}
\frac{\Gamma\ (n + \delta_{m}^{(s)}  - p + 1 - \imath\sigma )}
{\Gamma (j + \delta_{m}^{(s)}    - p + 2 - \imath\sigma )}.
\end{eqnarray*}
As a result, we can rewrite relation (\ref{eq2-16}) as follows:
\begin{eqnarray*}
\frac{\pi R^{3}_0e^{-\pi \sigma}\left|C_{nj}^{(s)}\right|^{2}}{2^{2j +2\delta_{m}^{(s)}    + 2}}
\left[\Gamma\left(2j +\delta_{m}^{(s)}      + 2\right)\right]^{2}
\frac{n + j +2\delta_{m}^{(s)}    + 1}
{\Gamma\left(n + j +2\delta_{m}^{(s)}    + 1\right)} \frac{\Gamma\left(n +\delta_{m}^{(s)}      -\imath\sigma\right)}
{\Gamma\left(j +\delta_{m}^{(s)}   + 1 - \imath\sigma\right)} \times  \\ [2mm]
\times
\sum\limits_{p=0}^{n-j-1}\frac{\left(-n + j + 1\right)_{p}
\left(j +\delta_{m}^{(s)}     + 1 + \imath\sigma\right)_{p}}
{p! \left(1 - n - \delta_{m}^{(s)}    + \imath\sigma\right)_{p}} \frac{\left(-1\right)^{p}}{\Gamma\left(j + \delta_{m}^{(s)}   + p + 2 + i\sigma\right)
\Gamma\left(n +\delta_{m}^{(s)}    - p + 1 - \imath\sigma\right)}
 \times \\ [2mm]
\times
\Biggl[\frac{\Gamma\left(n - j - p - 1\right)}{\Gamma\left(-p\right)} +   \frac{ (n - j - p - 1)(j +\delta_{m}^{(s)}      + p + 1 + \imath\sigma )}
{n + j +2\delta_{m}^{(s)}     + 1}\frac{\Gamma\left(n - j - p - 1\right)}{\Gamma\left(1 - p\right)}\Biggr]
= 1.
\end{eqnarray*}
Taking into account that in the last expression the first term is non-zero only for $p = n - j -1$, and the second only for $p = 0$ and $p = n - j -1$, we obtain the expression for the normalization constant \eqref{normsphere}.

\section{Generalized MICZ-Kepler system on three-dimensional hyperboloid}
The three-dimensional two-sheeted hyperboloid (pseudosphere) of radius $R_0$ is the hypersurface in the four-dimensional Minkowski space ($\mathbf{R}^{1.3},\;dy_0 dy_0-d\mathbf{y} d\mathbf{y})$ defined by the constraint
$y_0^2-\mathbf{y}^2=R^2_0$, where $\mathbf{y}=(y_1,y_2, y_3)$.

This constraint can be resolved in stereographic  coordinates $\mathbf{r}=(x_1,x_2,x_3)$:
\be
\mathbf{y}= \frac{ \mathbf{r}}{1-\frac{\mathbf{r}^2}{4R^2_0}}, \quad \mathbf{y}_0= R_0\frac{1+\frac{\mathbf{r}^2}{4R^2_0} }{1-\frac{\mathbf{r}^2}{4R^2_0} },
\ee
and  in hyperspherical coordinates $\tau, \theta, \varphi$:
\begin{eqnarray}
y_{1} +\imath y_{2} = R_0\sinh\tau \sin\theta e^{\imath\varphi}, \qquad y_{3} = R_0\sinh\tau \cos\theta, \qquad y_{0} = R_0\cosh \tau,
\label{eq3-1}
\end{eqnarray}
where $\tau\in [0,\infty ), \theta \in [0, \pi), \varphi \in [0, 2\pi)$.

Respectively,
\begin{eqnarray}
 x_{1} + \imath x_{2} = \frac{2R_0 \sinh\tau \sin\theta e^{\imath\varphi}}{1+\cosh\tau}, \qquad x_{3} =\frac{2R_0\sinh\tau \cos\theta }{1+\cosh\tau}.
\label{eq2-2p}
\end{eqnarray}

The induced metric on the hyperboloid  then takes the form:
\be
ds^2=\left( dy_0 dy_0 -d\mathbf{y} d\mathbf{y}\right)\vert_{y_0^2-\mathbf{y}^2=R^2_0}=\frac{d\mathbf{r}d\mathbf{r}}{(1-\frac{\mathbf{r}^2}{4R_0^2})^2}=R^2_0\left(d\tau^2+\sinh^2\tau(d\theta^2 +\sin^2\theta d\varphi^2)\right).
\ee
Thus, in stereographic  coordinates the metric takes the conformally flat form \eqref{cf} with
\be
g(r)=\left(1-\frac{\mathbf{r}^2}{4R_0^2}\right)^{-2} .
\label{gH}\ee
In these coordinates the rotational momentum generators $\mathbf{J}$ are as in the flat case \eqref{JI}.

The Coulomb potential on the three-dimensional hyperboloid has the form \cite{Schrod}:
\begin{eqnarray}
V  =-\frac{e^2}{R_0}\left(\frac{y_0}{|\mathbf{y}|} -1\right)= - \left(1-  \frac{\mathbf{r}^2}{4R^2_0}\right) \frac{e^2}{|\mathbf{r}|} +\frac{e^2}{R_0}=
  - \frac{e^{2}}{R_0}\left(\coth \tau - 1\right).
\label{eq3-2}
\end{eqnarray}

In analogy with the previous section, we define the hyperbolic analog of the generalized MICZ-Kepler potential \eqref{gCoulomb} by the expression:
\be
V_{gMICZ}= \frac{1}{g(r)}\left(\frac{\hbar^2s^2}{2\mu r^2} + \frac{\lambda_1}{r(r+x_3)}+ \frac{\lambda_2}{r(r-x_3)}\right)  - \left({1-\frac{\mathbf{r}^2}{4R_0^2}}\right)\frac{e^2}{ {r}}+\frac{e^2}{R_0},
\label{vh}\ee
 where  $g(r)$ is defined by \eqref{gH}.

The Hamiltonian of this system is given by \eqref{hamilt}, with  $g(r)$ and $V_{gMICZ}$ given by \eqref{gH} and \eqref{vh} respectively. 

Taking into account the spectral problem \eqref{sp}, we can represent the quasi-radial Schr\"odinger equation for the generalized MICZ-Kepler system on the hyperboloid as follows:
\begin{eqnarray}
\frac{1}{\sinh^{2}\tau}\frac{d}{d \tau}
\left(\sinh^{2}\tau \frac{d \mathcal{R}}{d \tau}\right) +
\Biggl[\frac{2\mu R^{2}_0}{\hbar^{2}}\left(E - \frac{e^{2}}{R_0}\right) - \frac{(j+ \delta_{m}^{(s)})(j+ \delta_{m}^{(s)} +1)}{\sinh^{2}\tau}
 + \frac{2\mu e^{2}R_0}{\hbar^{2}}\coth\tau\Biggr]
\mathcal{R}(\tau ) = 0. \nonumber
\end{eqnarray}
This equation has been solved in \cite{KMP}.
Using the solution presented there we obtain the discrete energy spectrum of the generalized MICZ-Kepler system on the three-dimensional hyperboloid:
\begin{eqnarray}
E_{n,m} = -\frac{\hbar^{2}}{2\mu R^{2}_0}\left[\left(n +  \delta_{m}^{(s)}    \right)^{2}
- 1\right] - \frac{\mu e^{4}}{2\hbar^{2}\left(n + \delta_{m}^{(s)} \right)^{2}}
+ \frac{e^{2}}{R_0},\qquad n = |s|+1, |s|+2, \ldots, n_{\text{max}}.
\label{eq3-10}
\end{eqnarray}
The discrete energy spectrum exists only under the condition 
\be 
0 \leq n \leq \left[\sigma - \delta_{m}^{(s)} - 1\right],
\label{n}\ee
with  $\left[a\right]$ denoting the integer part of $a$, and with 
 the parameter $\sigma$ defined as in the spherical case \eqref{sigma}
\be
\sigma = \frac{R_0}{r_0 (n+\delta^{(s)}_m)} =
 \frac{\mu e^2 R_0}{\hbar^2 (n+\delta^{(s)}_m)}.
\nonumber\ee

The condition \eqref{n} fixes the following 
maximal value of the quantum number $n$  
\be
n_{\text{max}} = \left[ \sqrt{\frac{\mu e^2 R_0}{\hbar^2}} - \delta_m^{(s)} \right],\qquad n_{\text{max}}\geq |s|+1,
\label{eq:bound}
\ee
 So, the discrete spectrum exists only for negative energies. The range of validity of $j,m$ is defined by \eqref{17}.

The quasi-radial wave function of the generalized MICZ-Kepler system is as follows
\begin{eqnarray}
\mathcal{R}_{nj}\left(\tau \right) = A_{nj}^{(s)}
\left(\sinh\tau\right)^{j + \delta_{m}^{(s)}   }
e^{\tau (n - j - \sigma - 1)}
 {_{2}F}_{1}\left(-n + j + 1, j  + \sigma+ \delta_{m}^{(s)}  + 1;
2j +2\delta_{m}^{(s)}  + 2; 1 - e^{-2\tau}\right),
\label{eq3-11}\end{eqnarray}
where $A_{nj}^{(s)} $ is the normalization constant.

It can be calculated from the requirement that the quasi-radial wave function \eqref{eq3-11} satisfies the condition:
\begin{eqnarray}
R^{3}_0\int\limits_{0}^{\infty} \sinh^{2}\tau \left|\mathcal{R}_{nj}\left(\tau \right)\right|^{2}d\tau = 1.
\label{eq3-12}
\end{eqnarray}
The calculation of the integral (\ref{eq3-12}) can be carried out in a similar way to the calculation of the normalization constant on the three-dimensional sphere. Namely, using relation (\ref{eq2-13}) we pass to hypergeometric functions of the argument $e^{-2\tau}$, perform integration in accordance with the formula \cite{Bateman1}:
\begin{eqnarray*}
\int\limits_{0}^{\infty} e^{-2\alpha t} \left(\sinh \beta t\right)^{\gamma}dt =
\beta^{-1}2^{-1-\gamma}B\left(\frac{\alpha}{2\beta}, 1 + \gamma\right),
\end{eqnarray*}
apply twice the relation (\ref{eq2-17}), and use formulae (\ref{eq2-15}) and (\ref{eq2-17-1}) for the gamma function.
Then we finally obtain:
\begin{equation}
A_{nj}^{(s)}
 = \frac{ 2^{j + \delta_{m}^{(s)}+ 1}  }
{\Gamma (2j +2\delta_{m}^{(s)}    + 2 ) }
 \sqrt{
 \frac{ (\sigma^{2} -  (n +\delta_{m}^{(s)}   )^{2})
\Gamma(n + j + 2\delta_{m}^{(s)}  + 1)
\Gamma(j + \sigma + \delta_{m}^{(s)}  + 1)}
{R^{3}_0 (n +\delta_{m}^{(s)} )(n - j - 1)!
\Gamma(\sigma - j - \delta_{m}^{(s)}   )}}  . \label{eq3-13}
\end{equation}
For $s = \lambda_{1} = \lambda_{2} = 0$, the energy spectrum \eqref{eq3-10} transforms into the energy spectrum of the Coulomb problem on hyperboloid obtained in \cite{Infeld} by the factorization method proposed by Schr\"odinger \cite{Schrod}, and the normalized quasi-radial function of the generalized MICZ-Kepler system transforms into the quasi-radial wave function of the Kepler-Coulomb problem on the  hyperboloid \cite{Barut}.\\

In this Section we suggested the generalized MICZ-Kepler system on the three-dimensional hyperboloid and constructed the discrete part of its spectrum and the respective wavefunctions. Similar to the generalized MICZ-Kepler system on the three-dimensional sphere, they depend on only two quantum numbers. This indicates that the system could/should have an additional integral generalizing the third expression in \eqref{cInt}.  

\section{Discussion and outlook}

In this paper we proposed generalized MICZ-Kepler systems on the three-dimensional sphere and pseudosphere (upper sheet of the two-sheted hyperboloid) and found their spectra and normalized wavefunctions.
The obtained solutions have well-defined limits: for $s = \lambda_{1} = \lambda_{2} = 0$ they reduce to the corresponding formulae for the Kepler-Coulomb problem on the three-dimensional (pseudo)sphere.
In the flat limit $R_0 \to \infty$ ($\chi \sim r/R_0$ for the sphere and $\tau \sim r/R_0$ for the pseudosphere) they reduce to the corresponding formulae for the generalized MICZ-Kepler system in three-dimensional Euclidean space.\\

The proposed systems were defined by the potential \eqref{gCoulombSphere}, with the special choice of $g(r)$ and $V(r)$ given by \eqref{gV}.
However, from our consideration it becomes clear that \eqref{gCoulombSphere} defines the relevant ``generalized MICZ-extension'' not only for the Coulomb systems on the three-dimensional sphere and hyperboloid, but also for a particle moving on any $so(3)$-invariant space (equipped with the metric \eqref{cf}) in the presence of any central potential $V(r)$. 
Indeed, substituting \eqref{gCoulombSphere} in \eqref{hamilt} we get
\be
\widehat{\mathcal{H}}= -\frac{\hbar^2}{2\mu}\triangle_{s}+\frac{1}{g(r)} \left(\frac{\hbar^2s^2}{2\mu r^2}+
 \frac{\lambda_{1}}{r(r+x_3)} +\frac{\lambda_{2}}{r(r-x_3)}\right)+V(r) =-\frac{\hbar^2}{2\mu} \triangle_r +\frac{\widehat{M}^{(s)}}{2\mu g(r)r^2} +V(r),
\label{hamilt2}
\ee
where $\triangle_{s}$,  $\triangle_r$  are defined by \eqref{hamiltR}, and $\widehat{M}^{(s)}$ is defined by \eqref{cInt}.
Passing to spherical variables  $x_3=r\cos\theta$, $x_1+\imath x_2 =r\sin\theta {\rm e}^{\imath \varphi}$ we  express $\widehat{M}^{(s)}$ via $\theta,\varphi$ only,  \eqref{eq17}. Then, taking into account \eqref{eq16} we reduce the 
Schr\"odinger equation $\widehat{H}\psi=E\psi$  to the radial  one.\\

The spectra of the proposed systems depend on two quantum numbers, which indicates that these systems should be ``minimally superintegrable'' (i.e., have, in addition to the Hamiltonian $\widehat{\mathcal{H}}^{\varepsilon}_{gMICZ}$, three functionally independent integrals of motion).  Two integrals,    $\widehat{J}_3$,  and $\widehat{{M}}^{(s)}_{gMICZ}$ , are defined in \eqref{cInt}, where the rotational momentum generators $\mathbf{J}$ are given by \eqref{JI}. The integral of motion generalizing the third expression in \eqref{cInt} 
needs to be found. If so,  perturbation of generalized (pseudo)spherical systems by an appropriate potential, e.g., the analog of the Stark term \cite{Vahagn}, will preserve the integrability and the separability of variables.  We hope to study this problem in forthcoming work.
 
We expect that the generalized MICZ-Kepler systems on the three-dimensional hyperboloid can be obtained by the Kustaanheimo-Stiefel transformation from the $\mathbf{CP}^2$-Rosochatius system \cite{INS}, in the spirit of \cite{conifold}.
Then, with an appropriate ``Wick rotation'' we could obtain the generalized MICZ-Kepler systems on the three-dimensional sphere. In this way we could, seemingly, obtain the fourth integral of the system, as well as its complete symmetry algebra. \\

In a similar way one can construct the five-dimensional counterpart of this system: the generalized Yang-Coulomb (or $SU(2)$ MICZ-Kepler) system on the five-dimensional sphere/hyperboloid, cf \cite{mny}.
 We expect that it could be constructed by an appropriate $SU(2)$-reduction from the analog of the Rosochatius system on the two-dimensional quaternionic projective space $\mathbb{HP}^2$.\\
 
 Finally, we suggest that the generalized MICZ-Kepler systems on curved spaces may find applications in the study of ring-shaped molecules (e.g. benzene-like structures) deposited on curved substrates (such as fullerenes and  negatively curved carbon allotropes), quantum dots on the  curved surfaces with magnetic monopole fields, as well as  and in the study of integrable systems on non-Euclidean geometries, which are of interests in the context of quantum optics and gravitational analogues. 

\acknowledgments
This work was partially supported by the Armenian State Committee of Higher Education and Science, project 21AG-1C062 (A.N.).


\begin{thebibliography}{99}

\bibitem{Zw}
D.~Zwanziger,
``Exactly soluble nonrelativistic model of particles with both electric and magnetic charges,''
Phys. Rev. \textbf{176} (1968), 1480-1488.

\bibitem{MIC}
H.~V.~McIntosh and A.~Cisneros,
``Degeneracy in the presence of a magnetic monopole,''
J. Math. Phys. \textbf{11} (1970), 896-916.

\bibitem{KS}
P.~Kustaanheimo and E.~Stiefel, ``Perturbation theory of Kepler motion based on spinor regularization,'' J. Reine Angew. Math. \textbf{218} (1965), 204-219.

\bibitem{iwai}
A.~Nersessian and V.~Ter-Antonian,
``'Charge-dyon' system as the reduced oscillator,''
Mod.\ Phys.\ Lett. A \textbf{9}(1994), 2431-2436.

V.~M.~Ter-Antonian and A.~Nersessian,
``Quantum oscillator and a bound system of two dyons,''
Mod.\ Phys.\ Lett.\ A \textbf{10}(1995), 2633-2638.

\bibitem{Higgs}
P.~W.~Higgs,
``Dynamical Symmetries in a Spherical Geometry. 1,''
J.\ Phys.\ A \textbf{12}(1979), 309-323.

H.~I.~Leemon,
``Dynamical Symmetries in a Spherical Geometry. 2,''
J. Phys. A \textbf{12}(1979), 489.

\bibitem{NP}
A.~Nersessian and G.~Pogosyan,
``On the relation of the oscillator and Coulomb systems on (pseudo)spheres,''
Phys. Rev. A \textbf{63}(2001), 020103(R).

V.~V.~Gritsev, Yu.~A.~Kurochkin and V.~S.~Otchik, ``Nonlinear symmetry algebra of the MIC-Kepler problem on the sphere $S^3$'', J. Phys. A \textbf{33}(2000), 4903.

\bibitem{mny}
L.~Mardoyan, A.~Nersessian and A.~Yeranyan,
``Relationship between quantum mechanics with and without monopoles,''
Phys. Lett. A \textbf{366}(2007), 30-35.

\bibitem{KNO}
S.~Krivonos, A.~Nersessian and V.~Ohanyan,
``Multi-center MICZ-Kepler system, supersymmetry and integrability,''
Phys. Rev. D \textbf{75}(2007), 085002.

A.~Nersessian and V.~Ohanyan,
``Multi-center MICZ-Kepler systems,''
Theor. Math. Phys. \textbf{155}(2008), 618-626.

\bibitem{Oz}
L.~Mardoyan, A.~Nersessian and M.~Petrosyan,
``The Stark effect in the charge dyon system,''
Theor. Math. Phys. \textbf{140}(2004), 958-964.

S.~Bellucci and V.~Ohanyan,
``Two-center quantum MICZ-Kepler system and Zeeman effect in the charge-dyon system,''
Phys. Lett. A \textbf{372}(2008), 5765-5772.

\bibitem{Vahagn}
A.~Nersessian and V.~Yeghikyan,
``Anisotropic inharmonic Higgs oscillator and related (MICZ-)Kepler-like systems,''
J. Phys. A \textbf{41}(2008), 155203.

S.~Bellucci and V.~Yeghikyan,
``The Coulomb problem on a 3-sphere and Heun polynomials,''
J. Math. Phys. \textbf{54}(2013), 082103.

\bibitem{smor}
J.~Fris, V.~Mandrosov, Ya.~A.~Smorodinsky, M.~Uhlir and P.~Winternitz, ``On higher symmetries in quantum mechanics'', Phys. Lett. \textbf{16}(1965), 354-356.

\bibitem{evans}
N.~W.~Evans, ``Super-integrability of the Winternitz system'', Phys. Lett. A \textbf{147} (1990), 483-486;
``Superintegrability in classical mechanics,''
Phys. Rev. A \textbf{41}(1990), 5666-5676.

A.~Guha and S.~Mukherjee, ``Exact solution of the Schr\"odinger equation with noncentral parabolic potentials'', J. Math. Phys. \textbf{28} (1987), 840-843.

Gh.~E.~Draganascu, C.~Campigotto, M.~Kibler, ``On a generalized Aharonov-Bohm plus Coulomb system'', Phys. Lett. A \textbf{170} (1992), 339-343.

M.~Kibler, L.~G.~Mardoyan, G.S.Pogosyan, ``On a generalized Kepler-Coulomb system: interbasis expansions'', Int. J. Quantum Chem. \textbf{52} (1994), 1301-1316.

\bibitem{Hartmann}
H.~Hartmann, ``Die Bewegung eines K\"orpers in einem ringf\"ormigen Potentialfeld'', Theor. Chim. Acta \textbf{24} (1972), 201-206.

H.~Hartmann, R.~Schuch, J.~Radke, ``Die diamagnetische Suszeptibilit\"at eines nicht kugelsymmetrischen Systems'', Theor. Chim. Acta \textbf{42} (1976), 1-3.

H.~Hartmann, R.~Schuch, ``Spin-orbit coupling for the motion of a particle in a ring-shaped potential'', Int. J. Quantum Chem. \textbf{18} (1980), 125-141.

\bibitem{Hothers}
M.~Kibler and T.~N\'{e}gadi, ``Motion of a particle in a ring-shaped potential: An approach via a nonbijective canonical transformation'', Int. J. Quantum Chem. \textbf{26} (1984), 405-410.

M.~Kibler and P.~Winternitz, ``Dynamical invariance algebra of the Hartmann potential'', J. Phys. A \textbf{20} (1987), 4097-4108.

C.~Quesne, ``A new ring-shaped potential and its dynamical invariance algebra'', J. Phys. A \textbf{21} (1988), 3093.

I.V.~Lutsenko, G.S.~Pogosyan, A.N.~Sissakian, and V.M.~Ter-Antonyan, ``Hydrogen atom as indicator of hidden symmetry of a ring-shaped potential'', Theor. Math. Phys. \textbf{83} (1990), 633-639.

A.S.~Zhedanov, ``Hidden symmetry algebra and overlap coefficients for two ring-shaped potentials'', J. Phys. A \textbf{26} (1993), 4633-4642.

\bibitem{Mardoyan1}
L.~Mardoyan,
``The Generalized MIC-Kepler system,''
J. Math. Phys. \textbf{44} (2003), 4981-4987;
``Spheroidal analysis of the generalized MIC-Kepler system,''
Phys. Atom. Nucl. \textbf{68} (2005), 1746-1755.

\bibitem{gMICZ}
P.~R.~Giri,
``Self-Adjointness of Generalized MIC-Kepler System,''
Mod. Phys. Lett. A \textbf{22} (2007), 2365-2377;
``Supersymmetric quantum mechanical generalized MIC-Kepler system,''
Mod. Phys. Lett. A \textbf{23} (2008), 895-904; ``Conformal anomaly in non-hermitian quantum mechanics,''
Int. J. Mod. Phys. A \textbf{25} (2010), 155-161.

I.~Marquette,
``Generalized MICZ-Kepler system, duality, polynomial and deformed oscillator algebras,''
J. Math. Phys. \textbf{51} (2010), 102105;
``Generalized Kaluza-Klein monopole, quadratic algebras and ladder operators,''
J. Phys. A \textbf{44} (2011), 235203.

M.~F.~Hoque, I.~Marquette and Y.~Z.~Zhang,
``A new family of $N$-dimensional superintegrable double singular oscillators and quadratic algebra $Q(3) \bigoplus so(n) \bigoplus so(N-n)$,''
J. Phys. A \textbf{48} (2015) no.44, 445207;
``Quadratic algebra for superintegrable monopole system in a Taub-NUT space,''
J. Math. Phys. \textbf{57} (2016) no.9, 092104.

M.~Salazar-Ram{\'\i}rez, D.~Ojeda-Guill{\'e}n and R.~D.~Mota,
``The number radial coherent states for the generalized MICZ-Kepler problem,''
J. Math. Phys. \textbf{57} (2016) no.2, 021704.

M.~Salazar-Ram{\'\i}rez, J.~A.~Mart{\'\i}nez-Nu{\~n}o and Cordero-L{\'o}pez,
``SU(1,1) coherent states for the Dunkl-Klein-Gordon equation in its canonical form,''
Few Body Syst. \textbf{67} (2026) no.1, 3

N.~\"{U}nal, ``Parametric-time coherent states for the generalized MIC-Kepler system'', J. Math. Phys. \textbf{47} (2006), 122105.
 
A.~Lavrenov,
``Generalized KS transformations, $ND$ singular oscillator and generalized MICZ-Kepler system,''
[arXiv:1908.03572 [math-ph]].

\bibitem{Dirac}
P.~A.~M.~Dirac, ``Quantised singularities in the electromagnetic field'', Proc. R. Soc. Lond. A \textbf{133} (1931), 60-72.

\bibitem{Mardoyan3}
L.~G.~Mardoyan and M.~G.~Petrosyan,
``4D singular oscillator and generalized MIC-Kepler system,''
Phys. Atom. Nucl. \textbf{70} (2007), 572-575.

M.~Petrosyan,
``Four-dimensional double singular oscillator,''
Phys. Atom. Nucl. \textbf{71} (2008), 1094-1101.

H.~Shmavonyan,
``$\mathbf{C}^N$-Smorodinsky{\textendash}Winternitz system in a constant magnetic field,''
Phys. Lett. A \textbf{383} (2019), 1223-1228.

\bibitem{Mardoyan4}
L.~G.~Mardoyan,
``Superintegrability and Coulomb-Oscillator Duality,''  Phys. Part. Nucl. {\bf 57} (2026)
No.1, 
41-143
[arXiv:2411.07733 [math-ph]].

\bibitem{Schrod}
E.~Schr\"odinger, ``A Method of Determining Quantum-Mechanical Eigenvalues and Eigenfunctions'', Proc. R. Irish Acad. \textbf{46} (1940), 9-16; ``Further Studies on Solving Eigenvalue Problems by Factorization'', Proc. R. Irish Acad. \textbf{46} (1940), 183-206; ``The Factorization of the Hypergeometric Equation'', Proc. R. Irish Acad. \textbf{47} (1941), 53-54.

\bibitem{KMP}
E.~G.~Kalnins, W.~Miller, Jr. and G.~S.~Pogosyan,
``The Coulomb oscillator relation on n-dimensional spheres and hyperboloids,''
Phys. Atom. Nucl. \textbf{65} (2002), 1119-1127 [arXiv:math-ph/0210002].

\bibitem{Flugge}
S.~Fl\"ugge, Practical Quantum Mechanics, Vol. 1, Springer, 1974.

\bibitem{Bateman1}
H.~Bateman and A.~Erd\'elyi, Higher Transcendental Functions, Vol. 1, McGraw-Hill, 1953.

\bibitem{Bailey}
W.~N.~Bailey, Generalized Hypergeometric Series, Cambridge Tracts No.32, Cambridge University Press, Cambridge, 1935.

\bibitem{Stiv}
A.~F.~Stevenson, ``Note on the "Kepler Problem" in a Spherical Space, and the Factorization Method of Solving Eigenvalue Problems'', Phys. Rev. \textbf{59} (1941), 842-843.

\bibitem{Mardoyan5}
S.~I.~Vinitsky, L.~G.~Mardoyan, G.~S.~Pogosyan, A.~N.~Sissakian, T.~A.~Strizh, ``Hydrogen atom in curved space. Expansion over free solutions on the three-dimensional sphere'', Phys. At. Nucl. \textbf{56} (1993), 321-327.

\bibitem{Infeld}
L.~Infeld and A.~Schild, ``A Note on the Kepler Problem in a Space of Constant Negative Curvature'', Phys. Rev. \textbf{67} (1945), 121-122.

\bibitem{Barut}
A.~O.~Barut, A.~Inomata and G.~Junker, ``Path integral treatment of the hydrogen atom in a curved space of constant curvature. II. Hyperbolic space'', J. Phys. A \textbf{23} (1990), 1179.

\bibitem{INS}
E.~Ivanov, A.~Nersessian and H.~Shmavonyan,
``$\mathbb{CP}^N$-Rosochatius system, superintegrability, supersymmetry,''
Phys. Rev. D \textbf{99} (2019) no.8, 085007.

\bibitem{conifold}
S.~Bellucci, A.~Nersessian and A.~Yeranyan,
``Quantum mechanics model on K\"ahler conifold,''
Phys. Rev. D \textbf{70} (2004), 045006.

\end{thebibliography}
\end{document}